\documentclass[prb,aps,twocolumn]{revtex4-1}
\usepackage{bm}
\usepackage{amsmath}
\usepackage{amssymb} 
\usepackage{epsfig}
\usepackage{graphicx}
\usepackage{color}
\usepackage{xcolor}
\usepackage{ulem}
\usepackage{cancel}
\usepackage{hyperref}
\newcommand{\pderiv}[2]{\frac{\partial #1}{\partial #2}}
\renewcommand{\phi}{\varphi}
\renewcommand{\kappa}{\varkappa}

\renewcommand{\i}{\mathrm i}

\newcommand{\eps}{\varepsilon}

\newcommand{\aver}[1]{\left \langle #1 \right \rangle}

\begin{document}

\title{Edge photogalvanic effect caused by optical alignment of carrier momenta in 2D Dirac materials}

\author{M.\,V.\,Durnev}
\author{S.\,A.\,Tarasenko}

\affiliation{Ioffe Physical-Technical Institute, 194021 St.\,Petersburg, Russia}

\begin{abstract}
We show that the inter-band absorption of radiation in a 2D Dirac material leads to a direct electric current flowing at sample edges. 
The photocurrent originates from the momentum alignment of electrons and holes and is controlled by the radiation polarization.
We develop a microscopic theory of such an edge photogalvanic effect and calculate the photocurrent for gapped and gapless 2D Dirac 
materials, also in the presence of a static magnetic field which introduces additional imbalance between the electron and hole currents.
Further, we show that the photocurrent can be considerably multiplied in a ratchet-like structure with an array of narrow strips.
%The results suggest that the edge photogalvanic effect can be used for the detection of terahertz and infrared radiation and its polarization state.
\end{abstract}

\maketitle

\section{Introduction}

Photoelectric phenomena in 2D materials are the topic of active research in recent years aimed at the development of photodetectors and energy harvesting devices~\cite{Koppens2014}. Various mechanisms of photocurrent generation based on photothermoelectric~\cite{Xu2010, Cai2014, Castilla2019}, photovoltaic and bolometric~\cite{Freitag2013}, plasmonic~\cite{Vicarelli2012,Muraviev2013, Bandurin2018}, photon drag~\cite{Karch2010,Entin2010,Obraztsov2014}, ratchet and photogalvanic~\cite{Tarasenko2011,Drexler2013, Geller2015, Fateev2017, Kheirabadi2018,Quereda2018} effects have been reported and studied. 
% second harmonic generation~\cite{Dean2009, Wehling2015}. 
%The above mentioned phenomena lead to generation of the direct electric current or electric current oscillating at the double frequency, which magnitude is quadratic in electric field of the incident radiation. 
Direct electric currents arise here due to the lack of space inversion in the 2D structure introduced by $p$-$n$ junctions, metallic contacts, substrate,
inhomogeneity of illumination or heating, or photon wave vector. Space inversion symmetry is also broken naturally at the edges of 2D material.
This gives rise to the edge photogalvanic effect (EPGE) observed recently in monolayer and bilayer graphene illuminated by THz radiation ~\cite{Karch2011, Candussio2020} and in graphene in the quantum Hall effect regime~\cite{Plank2018}. The EPGE is 
reminiscent of the surface photogalvanic effect studied in bulk semiconductor crystals and metal films~\cite{Magarill1979, Alperovich1981, Alperovich1989, Gurevich1993, Schmidt2015,Mikheev2018}. 
 
The EPGE in 2D materials has been studied so far for low-frequency radiation inducing indirect intraband (Drude-like) optical transitions in doped structures~\cite{Karch2011, Candussio2020,Plank2018}. These transitions are quite weak since they can occur only if the scattering of carriers by static defects or phonons is involved. With increasing the frequency of radiation direct optical transitions between the valence and conduction bands come into play and the absorption dramatically rises.
Even in 2D monolayers, such as graphene or transition metal dichalcogenides (TMCD), the absorbance related to the interband optical transitions is known to reach few percent~\cite{Nair2008}. Accordingly, one can expect the enhanced photoresponse in this regime.

Here, we study the edge photogalvanic effect in 2D materials caused by direct interband optical transitions. The edge photocurrent emerges due to the alignment of charge carrier momenta by linearly polarized electromagnetic wave with a subsequent scattering of the carriers at the edge and consists of electron and hole contributions. 
%The mechanism of edge current formation is reminiscent of the surface photogalvanic effect studied at the surfaces of bulk semiconductor crystals and metal films~\cite{Magarill1979, Alperovich1981, Gurevich1993, Mikheev2018}. 
We develop a microscopic theory of the EPGE for a large class of 2D materials with the gapped or gapless Dirac-like energy spectrum, such as monolayer and bilayer graphene, monolayers of TMCD, HgTe/CdHgTe quantum wells with close-to-critical thickness, etc. We show that the current is controlled by the radiation polarization, its magnitude reaches 1~nA per W/cm$^2$ of the radiation intensity for a single edge and can be enhanced to 1~$\mu$A in a ratchet-like structure consisting of multiple narrow strips. We also investigate the effect of a static magnetic field applied normally to the 2D plane and show that the field modifies the polarization dependence of the currents and introduces additional imbalance between the electron and hole currents. Taking into account the important role of edge regions in micro- and nano-scale devices, we expect that the EPGE can determine the photoresponse of small-size devices and find applications in detectors of terahertz and infrared radiation and radiation polarization.

The paper is organized as follows. In Sec.~\ref{model} we describe the model of the edge photocurrent formation, calculate the rate of optical transitions and momentum alignment in 2D Dirac materials, and present the microscopic theory of the EPGE in a semi-infinite system. In Sec.~\ref{Sec_B} we study the effect of a static magnetic field on the photocurrent. In Sec.~\ref{strip} we calculate the photocurrent and its spatial distribution in a single strip and a strip structure. Results of the paper are summarized in Sec.~\ref{summary}.

\section{Model and theory} \label{model}

\begin{figure}[htpb]
\includegraphics[width=0.49\textwidth]{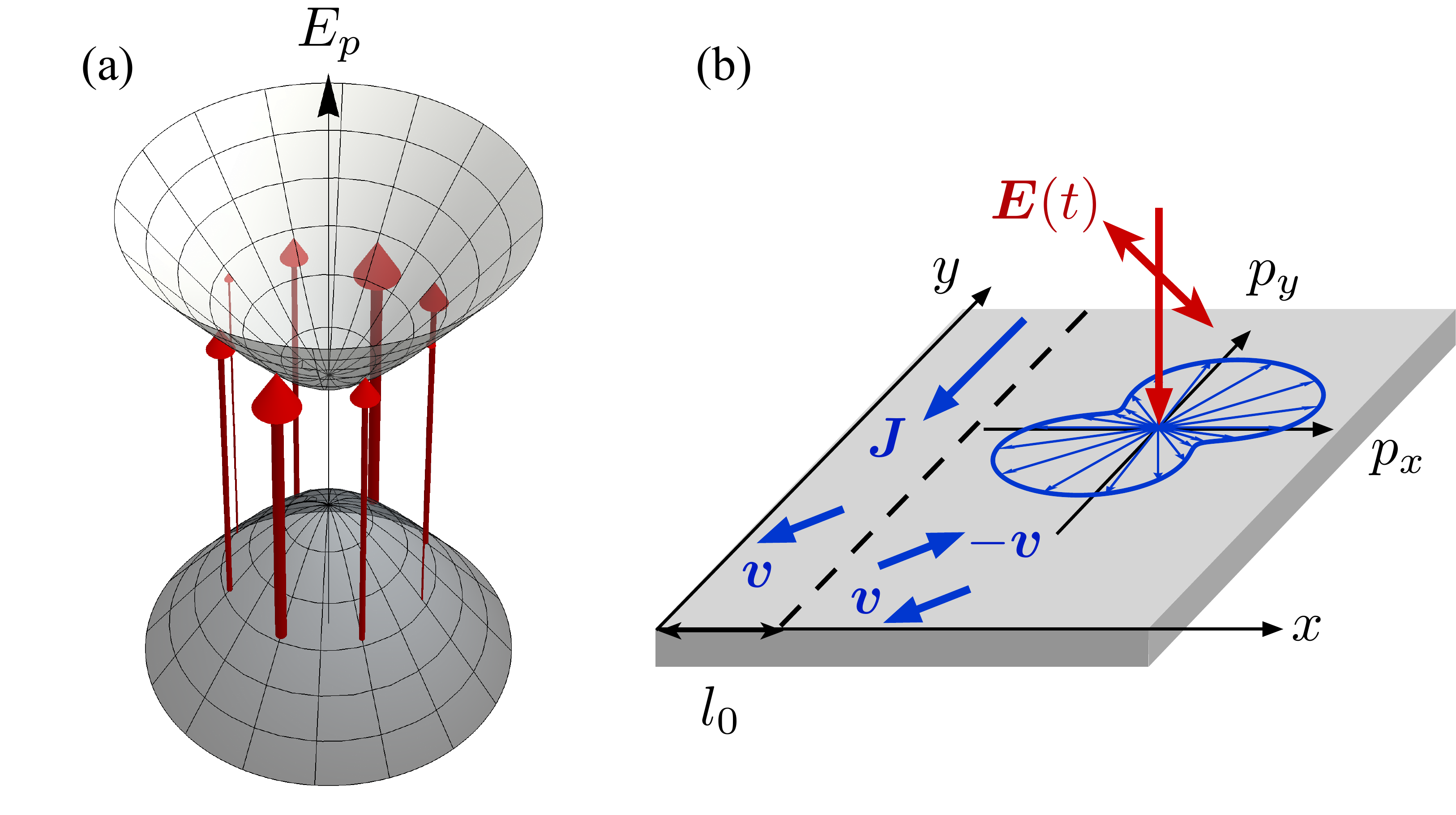}
\caption{\label{fig1} (a) Optical alignment of electron and hole momenta in 2D Dirac structure. 
The probability of interband transitions (shown by vertical arrows of different thicknesses) induced by linearly polarized radiation depends on the direction of the electron 
momentum $\bm p$. This leads to an anisotropic distribution of 
photoelectrons and photoholes in the $\bm p$ space. (b) Mechanism of the edge photogalvanic effect. Optical alignment of electrons 
(shown by closed blue curve) followed by electron scattering from the edge of 2D structure results in a direct electric current 
$\bm J$ flowing along the edge. The current flows within a narrow strip near the edge with the width determined by the mean free path 
$l_0$.      
}
\end{figure}

 The proposed microscopic mechanism of the edge photogalvanic effect is a two-step process sketched in Fig.~\ref{fig1}.
We consider that a semi-infinite 2D structure is illuminated by normally-incident linearly polarized radiation which causes 
direct optical transitions between the valence and conduction bands.  

At the first step, the absorption of linearly polarized radiation leads to the momentum alignment of photoelectrons (and photoholes), Fig.~\ref{fig1}a.
This phenomenon stems from the fact that the optical transition probability depends on the relative orientation between 
the quasi-momentum $\bm p$ and the electric field $\bm E$ of the radiation. The phenomenon is well known for 
bulk semiconductors~\cite{Alperovich1981,Zemskii1976, Dymnikov1976, Zakharchenya1982} and 2D systems based on semiconductor quantum wells~\cite{Merkulov1990}, graphene~\cite{Hartmann2011,Golub2011}, etc.
The optically aligned electrons and holes are characterized by anisotropic but even-in-$\bm p$ distribution in the momentum space 
(shown by closed blue curve in Fig.~\ref{fig1}b). Since the distribution is even in $\bm p$, no electric current is generated in the ``bulk'' of the 2D plane.

The direct electric current $\bm J$ emerges at the second step as a result of the scattering of optically aligned electrons 
at the edge, which introduces a local asymmetry in the electron distribution in the momentum space. Only those carriers who were created 
within the mean free path from the edge may contribute to the current, therefore the current flows in a narrow strip at the edge, as shown in Fig.~\ref{fig1}b. The edge current is expected to have distinct polarization dependence: it flows in the opposite directions for the radiation polarized at $\pm \pi/4$ angle with respect to the edge and vanishes if the radiation is polarized perpendicular to or along the edge.  

Below, we develop a microscopic theory of the edge photogalvanic effect. We calculate the excitation spectrum of the photocurrent
and its dependence on the band-structure parameters.

\subsection{Band structure and momentum alignment}

Consider first the band structure of 2D materials and interband optical transitions. 
The electron and hole states in a wide class of direct gap Dirac materials can be described by the effective Hamiltonian~\cite{Bernevig2006, Neto2009, Kormanyos2013}
\begin{equation}
H = a \, \bm{p} \cdot \bm{\sigma} + (\delta_0 + b p^2) \sigma_z + d p^2 I \,,
\end{equation}
where $\bm p=(p_x,p_y)$ is the electron momentum counted from the Dirac point, $\sigma_{j}$ ($j=x,y,z$) are the Pauli matrices, $I$ is the identity matrix, $\delta_0$, $a$, $b$, and $d$ are the band structure parameters.  The Hamiltonian contains all essential ingredients: 
the band gap $2 \delta_0$, the linear-in-$\bm p$ coupling between the valence- and conduction-band states, and the parabolic term $d p^2 I$ introducing electron-hole asymmetry. The Dirac states can have additional spin or/and valley degrees of freedom.

The energy spectrum and the wave functions of the conduction (``$c$'') and valence (``$v$'') bands in the electron representation are given by 
\begin{equation}
E_{c/v, p} = \pm \varepsilon_p + dp^2 
\end{equation}
and
\begin{equation}
\Psi_{c/v, \bm p} = \frac{1}{\sqrt{2\varepsilon_p (\varepsilon_p \mp \delta)}} 
\left(
\begin{array}{c}
-a p_- \\
\delta \mp \varepsilon_p
\end{array}
\right) ,
\end{equation}
respectively, where $\varepsilon_p = \sqrt{\delta^2 + (a p)^2}$, $\delta = \delta_0 + bp^2$, and  $p_{\pm} = p_x \pm i p_y$.

The Hamiltonian of electron-photon interaction has the form
\begin{equation}
V = - (e/c) \bm{A}(t) \cdot \nabla_{\bm p} H \,,
\end{equation}
where $e$ is the electron charge, $c$ is the speed of light, $\bm{A}(t) = \bm A \exp{(-i\omega t)} + {\rm c.c.}$ is the vector potential of the field, 
$\omega$ is its frequency, and $\bm A$ is its (complex) amplitude related to the electric field amplitude by $\bm{E}=(i\omega/c)\bm A$.  

The electromagnetic field excites electrons from the filled valence band to the empty conduction band. The matrix element of such optical transitions 
$(v, \bm p) \rightarrow (c, \bm p)$ has the form
\begin{equation}
\label{Vpp}
V_{\bm{p} \bm{p}} = - \frac{e |a|}{\varepsilon_p p c} 
\left[(b p^2 - \delta_0) (\bm A \cdot \bm p) - \i \eps_p (\bm A \times \bm p)_z \right]\:.
\end{equation}
The rate of optical transitions is given by Fermi's golden rule 
\begin{equation}\label{g_rate}
g_{\bm p} = \frac{2 \pi}{\hbar} |V_{\bm p \bm p}|^2 \delta(2 \varepsilon_p - \hbar\omega)\:,
\end{equation}
where $|V_{\bm p \bm p}|^2$ is found from Eq.~\eqref{Vpp} and has the form
\begin{eqnarray}\label{Vpp2}
|V_{\bm p \bm p}|^2 &=&  \frac{\pi e^2 a^2 I}{\varepsilon_p^2 \, \omega^2 c \, n} 
\{ (\delta_0 - bp^2)^2 +\varepsilon_p^2 + 2\varepsilon_p (\delta_0 - bp^2) S_3  \nonumber \\
&-& (a^2 + 4 b \delta_0 ) [(p_x^2 - p_y^2) S_1 + 2 p_x p_y S_2]  \} \,. 
\end{eqnarray}
Here, $I = c n |\bm E|^2/(2 \pi)$ is the intensity of radiation, $n$ is the refractive index of the dielectric medium surrounding the 2D Dirac material, and $S_1 = (|E_x|^2 - |E_y|^2)/|\bm E|^2$, $S_2 = (E_x E_y^* + E_y E_x^*)/|\bm E|^2$, and $S_3 = i(E_x E_y^* - E_y E_x^*)/|\bm E|^2$ are the Stokes parameters of the radiation polarization. 

The generation rate~\eqref{g_rate} includes the polarization-independent contribution, the terms $\propto S_1$ and $\propto S_2$ describing the momentum alignment of electrons by linearly polarized radiation, and the contribution sensitive to the circular polarization. The latter has the opposite signs for the pair of Dirac cones (e.g., spin subbands or valleys) related by time reversal symmetry and describes the spin/valley polarization of electrons by circularly polarized radiation~\cite{Mak2012}. In contract, the optical alignment of electron momenta by linearly polarized radiation is the same for all Dirac cones.    

\subsection{Edge photocurrent} \label{sec_zero_B}

In this section we calculate the dc edge current emerging in a semi-infinite sample at homogeneous illumination by linearly polarized radiation.
We consider a 2D Dirac material occupying a half-plane $x \geq 0$ with the edge parallel to $y$-axis, see Fig.~\ref{fig1}b.

The edge electric current consists of electron and hole contributions, $J_y^{e}$ and $J_y^{h}$, respectively, and is given by
\begin{equation}\label{J_total}
J_y = J_y^{e} + J_y^{h} \,, \;\;\; J_y^{e/h} = \int \limits_0^{+\infty}  j_y^{e/h}(x)  dx \,,
\end{equation}
where $j_y^{e}(x)$ and $j_y^{h}(x)$ are the local current densities. Below, we calculate the electron contribution $J_y^{e}$. 
The hole contribution $J_y^{h}$ can be computed in a similar way. 

The density of electric current in the conduction band is expressed via the electron distribution function 
$f(x, \bm p)$ as follows
\begin{equation}\label{j_y_def}
j_y^{e}(x) = e \nu \sum \limits_{\bm p} v_{y} f(x, \bm p)  \,,
\end{equation}
where $\nu$ is the factor of spin and valley degeneracy of Dirac states (e.g., $\nu = 4$ for graphene) and 
$\bm v = \nabla_{\bm p} E_{c,\bm p}$ is the electron velocity.
The distribution function is found from the kinetic equation
\begin{equation}\label{kinetic_eq}
v_{x} \pderiv{f}{x} = g_{\bm p} + {\rm St} f \,,
\end{equation}
where $g_{\bm p}$ is the optical generation rate of electrons and ${\rm St} f$ is the collision integral. 

The collision integral in the relaxation time approximation is given by
\begin{equation}
{\rm St} f = -\frac{f(x, \bm p) - \langle f(x, \bm p) \rangle }{\tau}  \,,
\end{equation}
where $\langle f(x, \bm p) \rangle$ is the distribution function averaged over the directions of $\bm p$ and $\tau$ is the relaxation time. Note that the collision integral above does not describe the relaxation of the zero angular harmonic of the distribution function. The corresponding relaxation times, governed by the processes of energy relaxation and recombination, are typically much larger than $\tau$ and do not affect the anisotropic part of the distribution function. 

The collision integral should be supplemented with the boundary condition at $x=0$. We consider {\it diffuse} or {\it specular} reflection of electrons from the sample edge. In the case of {\it diffuse} scattering, the distribution of the particles reflected from the edge is even in $p_y$, i.e., 
$ f(0, p_x >0, p_y) = f(0, p_x >0, -p_y)$. For {\it specular} reflection, the distribution satisfies $ f(0, p_x, p_y) = f(0, -p_x, p_y)$. 
Additional condition $\sum_{\bm p} v_x f(x, \bm p) = 0$ comes from the lack of electron flux to the edge in the absence of spatially inhomogeneous generation and recombination of carriers. 

To calculate the electric current along $y$ we decompose the distribution function $f(x, \bm p)$ into the symmetric and asymmetric in $p_y$ parts as follows 
\begin{equation}
f^{(s/a)}(x, p_x, p_y) = \frac{1}{2} [f(x, p_x, p_y) \pm f(x, p_x, -p_y)] \,.  
\end{equation}  
The asymmetric part satisfies the equation
\begin{equation}\label{eq_g_asym}
v_x \pderiv{f^{(a)}}{x} = - \frac{f^{(a)}}{\tau} + g_{\bm p}^{(a)} \,,
\end{equation}
where $g_{\bm p}^{(a)}$ is the asymmetric-in-$p_y$ part of the generation term $g_{\bm p}$. Solution of Eq.~\eqref{eq_g_asym} with the boundary conditions discussed above has the form
\begin{equation}\label{f_asym}
f^{(a)}(x,\bm p) = \tau g_{\bm p}^{(a)} + \tau \left[ \zeta g_{-p_x, p_y}^{(a)} - g_{\bm p}^{(a)} \right] \exp\left( - \frac{x}{v_x \tau} \right) \Theta (p_x)  \,,
\end{equation} 
where $\zeta$ is the dimensionless parameter defined by character of edge scattering ($\zeta = 0$ for {\it diffuse} scattering and $\zeta = 1$ for {\it specular} reflection) and $\Theta (p_x)$ is the Heaviside step function. 

Multiplying $f^{(a)}(x,\bm p)$ by $e v_y$, summing up the result over $\bm p$ and integrating by $x$, and taking into account that the generation rate $g_{\bm p}$ is an even function of $\bm p$, we obtain the electric current
\begin{equation}\label{J_electron}
J_y^{e} = - e \nu \frac{1+ \zeta_e}{2} \sum \limits_{\bm p} \tau_e^2 v_{e,x} v_{e,y} \, g_{\bm p} \,.
\end{equation} 
Here, the parameter of edge scattering specularity, the relaxation time, and the velocity related to the electrons in the conduction band 
are denoted as $\zeta_e$, $\tau_e$, and $\bm{v}_e$, respectively. We note that the result~\eqref{J_electron} in the case of {\it specular} reflection
can be also obtained without the explicit calculation of the distribution function~\eqref{f_asym} (see Appendix).

Similar calculations show that the electric current carried by holes in the valence hand has the opposite sign and is given by 
\begin{equation}\label{J_hole}
J_y^{h} = e \nu \frac{1+ \zeta_h}{2} \sum \limits_{\bm p} \tau_h^2 v_{h,x} v_{h,y} \, g_{\bm p} \,,
\end{equation} 
where $\zeta_h$, $\tau_h$, and $\bm{v}_h = - \nabla_{\bm p} E_{v, \bm p}$ are the corresponding hole parameters in the hole representation.

%\bl{In the case of diffuse reflection $f(x, p_x < 0, p_y)$ does not depend on $x$, and Eq.~\eqref{J_electron2} is simplified as follows
%\begin{equation}
%J_y^{e} =  - e \sum \limits_{\bm p} \tau_1 v_x v_y \Theta(v_x) f(\infty,\bm p) \,.
%\end{equation}
%Here we took into account that $\sum_{\bm p} \tau_1 v_x v_y \Theta(v_x) f(0,\bm p) = 0$ since $f(0,p_x>0, p_y)$ is an even function of $p_y$. If $g_{\bm p}$ is an odd function of $\bm p$, we obtain Eq.~\eqref{J_electron} with $\tau_2 \to \tau_1 \tau_2$.
%}

%Summing up the electron and hole contributions to the edge photocurrent we obtain
%%
%\begin{equation}\label{J_sum}
%J_y = \frac{e \nu}{2} \sum \limits_{\bm p} \left[ (1+ \zeta_h) \tau_h^2 v_{h,x} v_{h,y} - (1+ \zeta_e) \tau_e^2 v_{e,x} v_{e,y} \right] g_{\bm p} \,,
%\end{equation}
%%
%where $\eta_{j}$, $\tau_{j}$, and $\bm v_{j}$ ($j =e,h$) are the corresponding electron and hole parameters of edge scattering specularity, relaxation times, and velocities. 

Equations~\eqref{J_electron} and~\eqref{J_hole} are quite general and can be applied to any 2D material. They show that, in systems with full electron-hole symmetry, the net electric current $J_y^{e}+J_y^{h}$ vanishes. In real systems, where the electron-hole symmetry is lifted intrinsically (in the energy spectrum) or extrinsically (e.g., by doping), the net photocurrent is non-zero. 

For the interband optical transitions, the generation term is given by Eqs.~\eqref{g_rate} and~\eqref{Vpp2}.
Putting these equations into Eqs.~\eqref{J_electron} and~\eqref{J_hole}, summing up over $\bm p$, and assuming that $|b| \ll a^2/\delta_0$, 
we finally obtain the electron and hole contributions to the edge current  
\begin{multline}\label{J_eh} 
J_{y}^{e/h} = \pm \frac{e \eta a^2 (1+\zeta_{e/h}) \tau_{e/h}^2 v_{e/h}^2 \, p_*^2}{2 (\hbar \omega)^3}  \Theta(\hbar\omega - 2\delta_0) I S_2 ,
\end{multline} 
where $\eta = \pi \nu e^2 /(4 \hbar c n)$ is the absorbance of the 2D Dirac material at $\hbar\omega \gg \delta_0$,
$p_* = \sqrt{(\hbar\omega)^2-(2\delta_0)^2}  /(2a)$ is the momentum of photoexcited electrons and holes, 
$v_{e/h} = a \sqrt{1-(2 \delta_0 /\hbar\omega)^2} (1 \pm \hbar \omega d /a^2)$ are the corresponding electron and hole velocities,
and $\Theta(x)$ is the Heaviside step function. The relaxation times $\tau_e$ and $\tau_h$ are taken at the electron and hole energies 
$E_{e/h}(p_*) = \hbar \omega/2 \pm (d/4 a^2) [(\hbar\omega)^2-(2\delta_0)^2]$, respectively.

%
%
%\ST{
%%
%\begin{multline}\label{J_eh}
%J_{y}^{e/h} = \pm \frac{e \eta a^2}{8} \frac{[(\hbar\omega)^2 - (2\delta_0)^2]^2}{(\hbar \omega)^5} \left( 1 \pm \hbar\omega \frac{d}{a^2} \right)^2
%\\ \times (1+\zeta_{e/h}) \Theta(\hbar\omega - 2\delta_0) I S_2 ,
%\end{multline} 
%%
%where $\eta = \pi \nu e^2 /(4 \hbar c n)$ is the absorbance of the 2D Dirac material at $\hbar\omega \gg \delta_0$,
%$v_{e/h} = a \sqrt{1-(2 \delta_0 /\hbar\omega)^2} (1 \pm \hbar \omega d /a^2)$ are the velocities of the photoexcited electrons and holes, 
%and $\Theta(x)$ is the Heaviside step function. The relaxation times $\tau_e$ and $\tau_h$ are taken at the electron and hole energies 
%$E_{e/h} = \hbar \omega/2 \pm (d/4 a^2) [(\hbar\omega)^2-(2\delta_0)^2]$, respectively.
%}
%

%\ST{
%%
%\begin{multline}\label{J_eh}
%J_{y}^{e/h} = \pm \frac{e \eta }{8} \frac{(\hbar\omega)^2 - (2\delta_0)^2}{(\hbar \omega)^3}  \Theta(\hbar\omega - 2\delta_0)
%\\ \times (1+\zeta_{e/h}) \tau_{e/h}^2 v_{e/h}^2 I S_2 ,
%\end{multline} 
%%
%where $\eta = \pi \nu e^2 /(4 \hbar c n)$ is the absorbance of the 2D Dirac material at $\hbar\omega \gg \delta_0$,
%$v_{e/h} = a \sqrt{1-(2 \delta_0 /\hbar\omega)^2} (1 \pm \hbar \omega d /a^2)$ are the velocities of the photoexcited electrons and holes, 
%and $\Theta(x)$ is the Heaviside step function. The relaxation times $\tau_e$ and $\tau_h$ are taken at the electron and hole energies 
%$E_{e/h} = \hbar \omega/2 \pm (d/4 a^2) [(\hbar\omega)^2-(2\delta_0)^2]$, respectively.
%}

As follows from Eq.~\eqref{J_eh}, the polarization dependence of the edge current is determined by the Stokes parameter $S_2 = (E_x E_y^* + E_y E_x^*)/|\bm E|^2$.
The current reaches maxima of the opposite signs for the radiation polarized at $\pm 45^\circ$  with respect to the edge and vanishes for the radiation polarized along or normal to the edge, in accordance with the model sketched in Fig.~\ref{fig1}. 

\begin{figure}[htpb]
\includegraphics[width=0.47\textwidth]{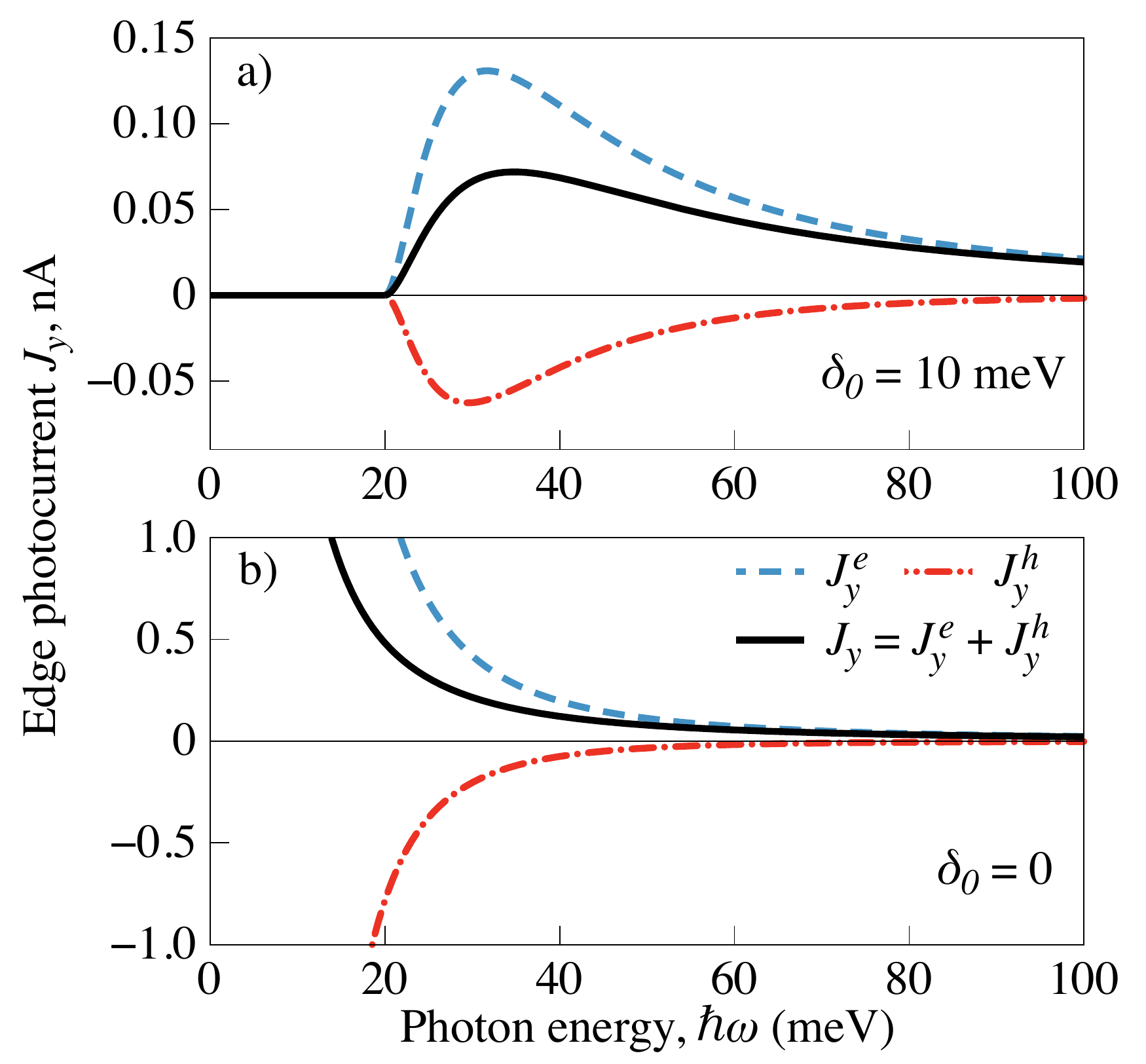} 
\caption{\label{fig_zeroB}
Excitation spectra of the electron $J_y^{e}$ and hole $J_y^{h}$ contributions to the edge photocurrent and the total edge photocurrent $J_y$ for gapped and gapless 2D Dirac materials. The spectra are calculated after Eq.~\eqref{J_eh} for short-range scattering of carriers in the 2D bulk and \textit{specular} scattering from the edge ($\zeta_{e/h} = 1$),  the band structures parameters $a = 10^8$~cm/s and $\delta_0 d/a^2 = 0.03$, the spin and valley degeneracy $\nu = 4$, the band gap $2 \delta_0 = 20$~meV, the relaxation time $\tau_0(10\,{\rm meV}) = 1$~ps, the refractive index of the surrounding medium $n = 3$, and the radiation intensity $I = 1$~W/cm$^2$. The radiation is linearly polarized at 45$^\circ$  with respect to the structure edge.
}
\end{figure}

Figure~\ref{fig_zeroB} shows the excitation spectra of the electron and hole currents, $J_y^e$ and $J_y^h$, respectively, and the total edge photocurrent $J_y = J_y^e + J_y^h$ in gapped and gapless systems. The spectra are calculated after Eq.~\eqref{J_eh} for the case when the energy dependence of the electron/hole scattering rate $\tau_{e/h}^{-1}(\varepsilon)$ follows the electron/hole density of states $D_{e/h}(\varepsilon)$, which is relevant to short-range scattering. For weak electron-hole asymmetry ($|d| \ll a^2/\delta_0$), the densities of states have the form
\[
D_{e/h} (\varepsilon) = \frac{\nu}{2\pi \hbar^2}\frac{\varepsilon}{a^2} \left[ 1 \mp \frac{d}{a^2} \frac{(3\varepsilon^2-\delta_0^2)}{\varepsilon} \right] \Theta(\varepsilon - \delta_0) \:,
\]
and we plot the spectra for the relaxation times 
\[
\tau_{e/h} (\varepsilon) = \tau_0(\tilde{\varepsilon}) \frac{\tilde{\varepsilon}}{\varepsilon} 
\left[ 1 \pm \frac{d}{a^2} \frac{(3\varepsilon^2-\delta_0^2)}{\varepsilon} \right]\:,
\]
where $\tau_0(\tilde{\varepsilon})$ is the relaxation time at the energy $\tilde{\varepsilon}$ in the absence of electron-hole asymmetry. 
The parameters used for calculations are given in the caption of Fig.~\ref{fig_zeroB}.  

In gapped Dirac materials (Fig.~\ref{fig_zeroB}a), the edge photocurrent is generated if the photon energy exceeds the band gap. Above this threshold, the photocurrent increases with the photon energy, reaches a maximum, and then decreases. In gapless materials with the Fermi energy lying at the Dirac point (Fig.~\ref{fig_zeroB}b), the photocurrent monotonically decreases with the photon energy. 
The electron and hole contributions to the photocurrent can be estimated as 
$J_y^{e/h} \sim e \eta  I l_{e/h}^2 / (\hbar\omega)$, where $l_{e/h} =  v_{e/h} \tau_{e/h}$ are the mean free paths of photoexcited electrons and holes.
For the mean free path of 1~$\mu$m and $\hbar\omega = 20$~meV, the photocurrent normalized by the radiation intensity is of the order of nA$\,$cm$^2$/W.
% corresponding to realistic HgTe/CdHgTe quantum well structures~\cite{Konig2007} $J_y^{e/h} \sim 1$~nA. 

\section{Effect of magnetic field \label{Sec_B}}

Now, we study the effect of a magnetic field $\bm B$ applied along the 2D plane normal $z$ on the edge photocurrent. We show that the magnetic field modifies the polarization dependence and magnitude of the photocurrent and, more interestingly, enables the generation of a net electric current even in a system with electron-hole symmetry. Similarly to the consideration in the previous section, we calculate the electron contribution to the edge current and then discuss both the electron and hole contributions. 

In a classical magnetic field, the kinetic equation for the steady-state distribution function of electrons in the conduction band has the form
\begin{equation}\label{kinetic_eq_B}
v_x \pderiv{f}{x} + \frac{e}{c} (\bm v \times \bm B) \cdot \pderiv{f}{\bm p}  = g_{\bm p} -\frac{f(x, \bm p) - \langle f(x, \bm p) \rangle }{\tau} \,.
\end{equation}
Multiplying Eq.~\eqref{kinetic_eq_B} by $v_y$, averaging the result over the direction of $\bm p$, and taking into account that 
$\langle v_y g_{\bm p} \rangle= 0$ we obtain the equation 
\begin{equation}\label{kin_eq_B_over}
\left\langle v_x v_y \pderiv{f}{x} \right\rangle + \frac{eB_z}{c} \left\langle v_x v_y \pderiv{f}{p_y} - v_y^2 \pderiv{f}{p_x} \right\rangle = 
- \frac{\langle v_y f \rangle}{\tau}  \,.
\end{equation}
Further, multiplying Eq.~\eqref{kin_eq_B_over} by $\tau$ and summing up over $\bm p$ we obtain after some algebra
\begin{equation}\label{kin_eq_B_sum}
j_y^e(x) = - e \nu \sum_{\bm p} \tau v_x v_y \pderiv{f}{x} + \frac{eB_z}{c} \sum_{\bm p} \frac{\tau}{m_c} v_x f \,,
\end{equation} 
where $m_c= p/v$ is the cyclotron mass. 

In the absence of spatially inhomogeneous recombination of carriers, the electron (and hole) flux to the edge given by the term $\aver{v_x f}$
vanishes. Then, the second term in the right-hand side of Eq.~\eqref{kin_eq_B_sum} can be neglected, and Eq.~\eqref{kin_eq_B_sum} yields 
\begin{equation}\label{J_electron_B}
J_y^e =  - e \nu \sum \limits_{\bm p} \tau v_x v_y [f(\infty,\bm p) - f(0,\bm p)] \,.
\end{equation}
%
%Equation~\eqref{J_electron_B} coincides formally with Eq.~\eqref{J_electron2}. However, $f(x,\bm p)$ here is the electron distribution function 
%in the presence of magnetic field.
The edge current is determined by the difference between the steady-state distribution functions at the edge and in the 2D bulk.

For \textit{specular} reflection of carriers at the edge, $f(0,\bm p)$ is even in $p_x$ and the term $\sum_{\bm p} \tau v_x v_y f(0,\bm p)$ vanishes.
The rest contribution $\sum_{\bm p} \tau v_x v_y f(\infty,\bm p)$ can be readily calculated analytically from the kinetic Eq.~\eqref{kinetic_eq_B} with the first term in the left-hand side being neglected. 
Such a calculation with the generation term~\eqref{g_rate} shows that the anisotropic part of the electron distribution function far from the edge has the form
\begin{eqnarray}
\delta f(\infty,\bm p) = - \frac{8 \pi^2 \tau e^2 a^2 (a^2 + 4b \delta_0) I}{(\hbar\omega)^3  \omega c n} 
\delta(2 \varepsilon_p - \hbar\omega)  \nonumber \\
\times \left[ (p_x^2 - p_y^2) \frac{S_1 + 2 \omega_c \tau S_2}{1+(2\omega_c\tau)^2} 
+ 2 p_x p_y \frac{S_2 - 2 \omega_c \tau S_1}{1+(2\omega_c\tau)^2} \right]  , \;\;
\end{eqnarray}
where $\omega_c = eB_z/{m_c c}$ is the cyclotron frequency. Finally, the electron and hole contributions to the edge photocurrent at 
$|b| \ll a^2/\delta_0$ are given by
\begin{multline}\label{J_final_B}
J_{y}^{e/h} = \pm \frac{e \eta a^2 \tau_{e/h}^2 v_{e,h}^2  \, p_*^2}{(\hbar\omega)^3} \frac{S_2  \mp 2\omega_{e/h} \tau_{e/h} S_1 }{1+(2\omega_{e/h} \tau_{e/h})^2} \\
\times \Theta(\hbar\omega - 2\delta_0)  I \,,
\end{multline}
where  $\omega_{e/h} = e B_z /(m_{e/h} c)$ are the cyclotron frequencies of the photoexcited electrons and holes and
$m_{e/h} = p_*/v_{e/h} = \hbar\omega/[2 a^2 (1 \pm \hbar \omega d /a^2)]$ are the corresponding effective masses.

%$p_{\rm tr} = (\hbar\omega/ 2a)\sqrt{1-(2\delta_0 / \hbar\omega)^2}$ is the momentum of photoexcited carriers, and
%$v_{e/h} = a \sqrt{1-(2\delta_0 / \hbar\omega)^2} (1 \pm \hbar \omega d /a^2)$ are the electron and hole velocities.

For {\it diffuse} scattering from the edge, the contribution to the edge current from the sum $\sum_{\bm p} \tau v_x v_y f(\infty,\bm p)$ is still given by Eq.~\eqref{J_final_B}. The second contribution in Eq.~\eqref{J_electron_B} given by the term $\sum_{\bm p} \tau v_x v_y f(0,\bm p)$ does not vanish any more and has to be calculated numerically.

\begin{figure}[htpb]
\includegraphics[width=0.48\textwidth]{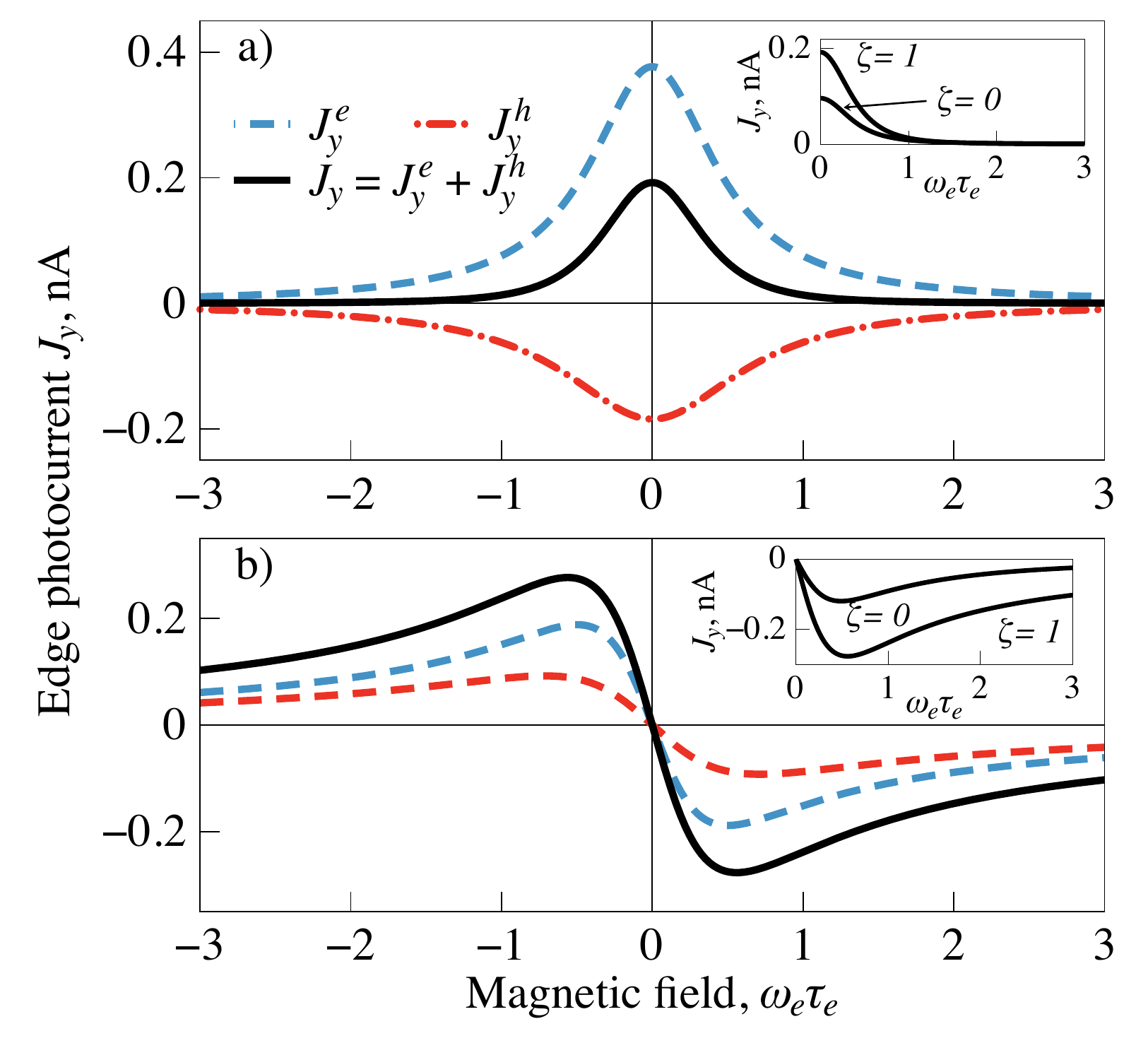}
\caption{\label{fig:fig3} 
Magnetic field dependence of the edge photocurrents excited by radiation (a) linearly polarized at 45$^\circ$ with respect to the structure edge
(the Stokes parameter $S_2 = 1$) and (b) linearly polarized along the edge (the Stokes parameter $S_1 = 1$). $J_y^e$ and $J_y^h$, and 
$J_y$ are the electron and hole contributions to the photocurrent, and the total photocurrent, respectively. The photocurrents are calculated for  \textit{specular} scattering from the edge, the band gap $2 \delta_0 = 20$~meV, the photon energy
$\hbar \omega = 30$~meV, the spin and valley degeneracy $\nu = 4$, the refractive index of the surrounding medium $n = 3$, the electron and hole mean free paths $l_e = 1$~$\mu$m and $l_h = 0.7~l_e$, respectively,  and the radiation intensity $I = 1$~W/cm$^2$.
The insets show the comparison between the total edge photocurrents in the structures with \textit{specular} ($\zeta_{e/h} = 1$) and \textit{diffuse} 
($\zeta_{e/h} = 0$) edge scattering.} 
\end{figure}

Figure~\ref{fig:fig3} shows the dependence of the edge photocurrent on magnetic field calculated after Eq.~\eqref{J_final_B} for {\it specular} reflection of carriers from the edge.
Figures~\ref{fig:fig3}a and~\ref{fig:fig3}b correspond to the radiation polarized at 45$^\circ$ to the edge and along the edge, respectively.
Since $\omega_{e/h} \tau_{e/h} \propto l_{e/h}$, where $l_{e/h} = v_{e/h} \tau_{e/h}$ are the mean free paths of the carriers, the difference between the hole and electron contributions to the photocurrent is determined by the single parameter $l_h/l_e$.  In the calculations we set $l_h/l_e = 0.7$. 
%To calculate $J_y$ for \textit{diffuse} scattering at the edge, we numerically solve Eq.~\eqref{kinetic_eq_B} to find $f(x, \bm p)$ and, subsequently, the contribution to the edge current given by the term  $\sum_{\bm p} \tau v_x v_y f(0,\bm p)$ in Eq.~\eqref{J_electron_B}.  

The radiation polarized at 45$^\circ$ to the edge induces the electron $J_y^{e}$ and hole $J_y^{h}$ edge photocurrents of the opposite sign, Fig.~\ref{fig:fig3}a. The photocurrents decrease in the magnetic field $B_z$ following the Hanle curves, see Eq.~\eqref{J_final_B}.
The net electric current is non-zero because of the electron-hole asymmetry.  

In the magnetic field, the edge photocurrents can be also excited by radiation polarized along (as well as perpendicular to) the edge, Fig.~\ref{fig:fig3}b.
For this polarization, the electron $J_y^{e}$ and hole $J_y^{h}$ contributions have the same sign and add up to each other. The net electric current
does not require the presence of electron-hole asymmetry in the 2D system. Therefore, quite a strong photoelectric response can be expected even in close-to-symmetric systems like graphene.

The \textit{diffuse} edge scattering leads to a reduction of the edge photocurrent as compared to the \textit{specular} one, as shown in the insets of Figs.~\ref{fig:fig3}a and~ \ref{fig:fig3}b. At zero magnetic field, the ratio of the edge currents in the structures with mirror ($\zeta_{e/h} = 1$) and rough ($\zeta_{e/h} = 0$) edges is 2:1, as also follows from Eq.~\eqref{J_eh}, whereas at $B_z \neq 0$ the ratio depends on the magnetic field and the radiation polarization.

\begin{figure}[htpb]
\includegraphics[width=0.48\textwidth]{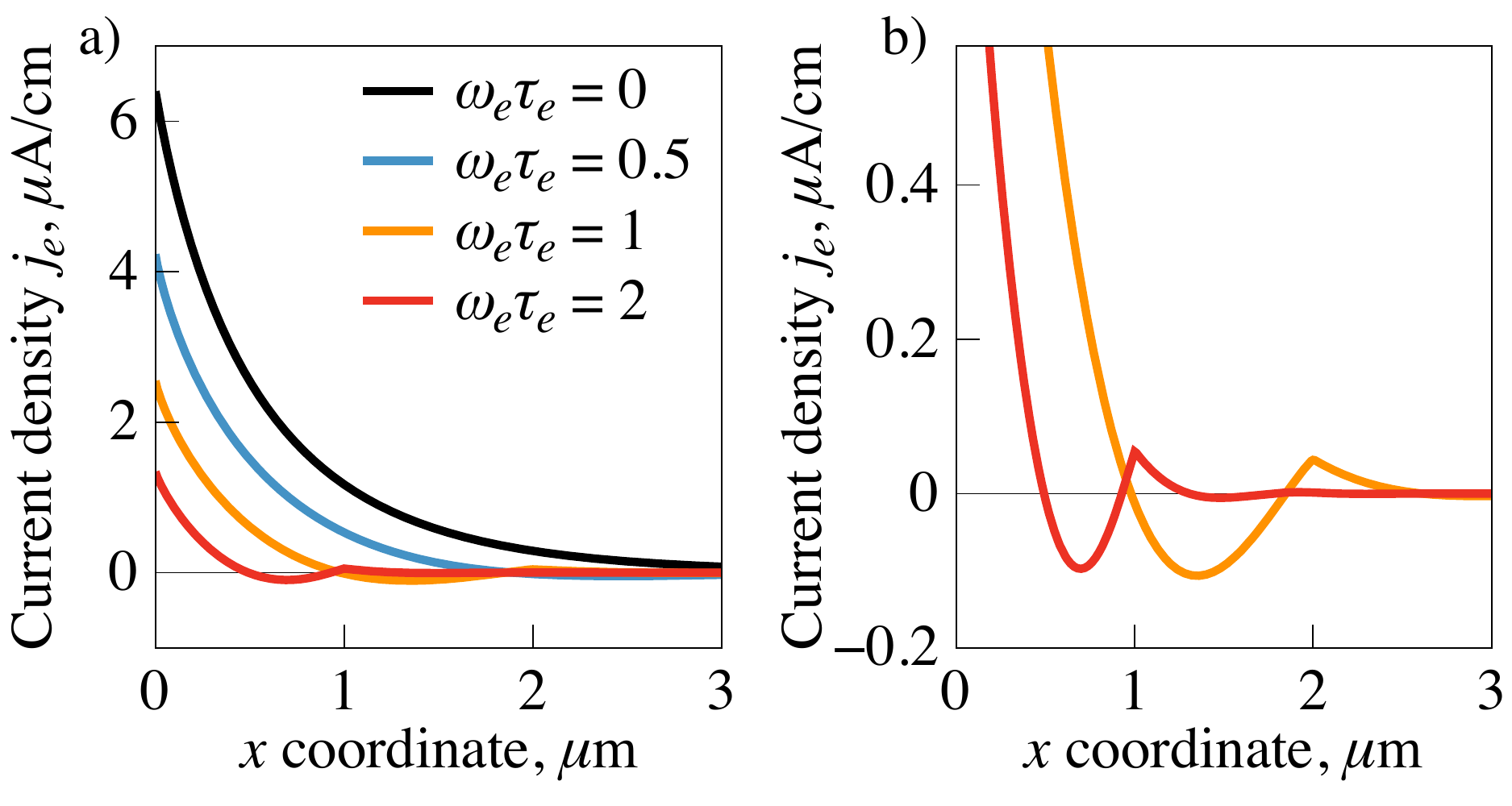}
\caption{\label{fig:fig4} 
Spatial distributions of the electron current density $j_y^e(x)$ near the sample edge at different magnetic fields: (a) $\omega_e \tau_e = 0,~0.5,~1,~2$ and (b) $\omega_e \tau_e = 1,~2$ at larger scale. The distributions are calculated for {\it specular} reflection of electrons from the edge, the radiation with the Stokes parameter $S_2 = 1$, and the parameters listed in the caption of Fig.~\ref{fig:fig3}.
}
\end{figure}

Numerical solution of the kinetic Eq.~\eqref{kinetic_eq_B} allows us to find the spatial distribution of the photocurrent density $j_y(x)$. The results of such calculations are shown in Fig.~\ref{fig:fig4}. As expected, the edge photocurrent flows within a narrow strip near the edge. 
The width of this strip is of the order of the carrier mean free path $l_{e/h}$ at zero magnetic field and close to the cyclotron diameter 
$2v_{e/h}/\omega_{e/h}$ at high magnetic fields. Interestingly, at $\omega_{e/h} \tau_{e/h} \gtrsim 1$, the edge photocurrent flows in the opposite directions in different regions near the edge, see Fig.~\eqref{fig:fig4}b.

In addition to the current controlled by the radiation polarization, in an external magnetic field there can emerge a polarization-independent 
photocurrent. Physically, it originates from the 2D analogue of the the Kikoin-Noskov photo-electro-magnetic effect known for 3D materials~\cite{Kikoin1934,Chernichkin2012}. The effect is related to an enhanced recombination rate of electrons and holes at the edge of the material, which may occur due to increased density of defect at the edge. The edge recombination leads to a local decrease of the electron and hole densities, which induces diffusive fluxes of the both types of carriers. At zero magnetic field, the fluxes are directed to the edge. 
The net electric current is zero because the currents carried by electrons and holes compensate each other in the steady-state regime. 
The flux density can be estimated as $i_x(x) \sim a_0 (\Delta \gamma_{\rm r} / \gamma_{\rm r}) G \exp(-\sqrt{\gamma_{\rm r}/D} \, x)$
where $a_0$ is a width of the order of few lattice constants where the recombination rate in enhanced, 
$\gamma_{\rm r} = 1/\tau_{\rm r}$, $\tau_{\rm r}$ is the lifetime of carriers far from the edge, $\Delta \gamma_{\rm r}$ is the extra recombination probability near the edge, $G = \eta(\omega) I/(\hbar\omega)$ is the rate of carrier generation, $\eta(\omega)$ is the absorbance, and $D$ is the diffusion coefficient of thermalized carriers. The external magnetic field deflects the fluxes in the opposite directions giving rise to a net electric current along the edge. This edge photocurrent flows within the diffusion length $\sqrt{D \tau_{\rm r}}$ near the edge and can be estimated as  
\begin{equation}\label{J_KN}
J_y^{\rm KN} \sim  \frac{e\eta a_0 \sqrt{D \tau_r}}{\hbar\omega} \frac{\Delta \gamma_{\rm r}}{\gamma_{\rm r}} 
\frac{\omega_c\tau}{1+(\omega_c\tau)^2} I \,.
\end{equation}
The ratio of the Kikoin-Noskov current~\eqref{J_KN} to the current caused by momentum alignment~\eqref{J_final_B} can be estimated as 
$(\Delta \gamma_{\rm r}/\gamma_{\rm r}) (a_0/l) \sqrt{\tau_{\rm r}/\tau}$, where $l = v \tau$ is the mean free path. One can expect that, 
in high-mobility structures, the current caused by momentum alignment predominates.

\section{Photocurrent in a strip structure \label{strip}} 

\begin{figure}[htpb]
\includegraphics[width=0.48\textwidth]{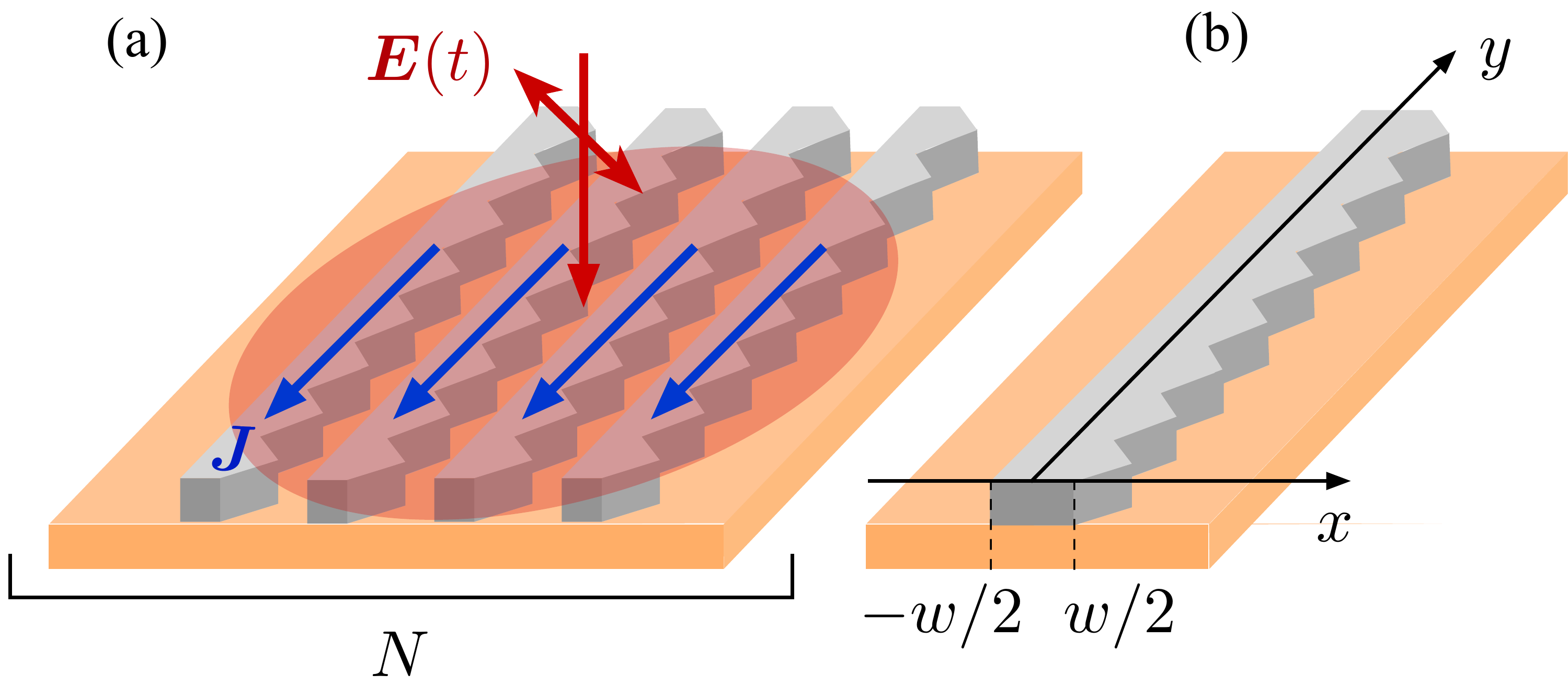}
\caption{\label{fig:fig5} 
(a) Sketch of a ratchet structure consisting of $N$ narrow asymmetric strips. The incident radiation induces dc electric current $\bm J$ in each strip, and the total current flowing in the structure is $N \bm J$. (b) Geometry of an individual strip.
}
\end{figure}

The edge photocurrent is formed in a narrow strip of the width determined by the carrier mean free path. The bulk of 2D system absorbs
radiation but is not involved in the current generation. Therefore, thinking of possible design of the structure with enhanced photoresponse we consider a ratchet structure consisting of $N$ narrow asymmetric strips, Fig~\ref{fig:fig5}a.  Optical excitation of such a structure leads to a photocurrent in each strip and the total current is increased by the factor of $N$.

Now, we calculate the photocurrent in an individual strip and study how it depends on the strip width and boundary conditions.
We consider that the 2D Dirac material is located at $-w/2 \leq x \leq w/2$, where $w$ is the strip width, Fig.~\ref{fig:fig5}b.
The distribution function of electrons in the strip is then a solution of the kinetic Eq.~\eqref{kinetic_eq_B} 
with boundary conditions at $x = -w/2$ and $x = w/2$. At zero magnetic field, the asymmetric-in-$p_y$ part of the electron distribution function 
has the form
\begin{equation}\label{f_strip}
f^{(a)}(x,p_x,p_y) = \tau \left[ g_{\bm p}^{(a)} + C_{\bm p} \exp\left( - \frac{x}{v_x \tau} \right) \right] \,,
\end{equation}
where 
\begin{eqnarray}
C_{\bm p} = &-& \frac{S(w/| 2v_x\tau |)}{S(w/|v_x \tau|)} g^{(a)}_{\bm p}  \\
&+&  [\zeta_l \Theta(p_x) + \zeta_r \Theta(-p_x)] \frac{\sinh(w/|2v_x\tau|)}{S(w/|v_x \tau|)} g^{(a)}_{-p_x,p_y} \nonumber \,,
\end{eqnarray}
$S(x)$ is the function defined by $S(x)= [\exp(x) - \zeta_l\zeta_r \exp(-x)]/2$, $g^{(a)}_{\bm p}$ is the asymmetric-in-$p_y$ part of $g_{\bm p}$,
and $\zeta_{l/r}$ are the parameters of scattering specularity at the left/right edges of the strip, respectively.    

Using the same method as described in Sec.~\ref{Sec_B}, one can show that the total electron photocurrent in the strip
$J_y^e = \int_{-w/2}^{w/2} j_e(x) dx$ is given by 
\begin{equation}
\label{Je_strip}
J_y^e =  - e \nu \sum \limits_{\bm p} \tau v_x v_y [f(w/2,\bm p) - f(-w/2,\bm p)] \, .
\end{equation}
The photocurrent is proportional to the difference of the distribution functions at the right and left edges of the strip and vanishes in a strip possessing the $x \to - x$ mirror symmetry. This symmetry is violated if the strip has an asymmetric shape, as shown in Fig.~\ref{fig:fig5}, or an asymmetric static potential $U(x)$. The current $J_y^e$ is also nonzero in a strip with different roughness of the left and right edges.

Substituting the distribution function Eq.~\eqref{f_strip} at $x = \pm w/2$ into Eq.~\eqref{Je_strip}, we finally obtain
\begin{equation}
\label{Je_strip_3}
J_y^e =  e \nu \left( \zeta_r - \zeta_l \right) \sum \limits_{\bm p} \tau^2 v_x v_y \frac{[\exp(w/v_x\tau)- 1]^2 g_{\bm p}}{\exp(2w/v_x\tau) - \zeta_l \zeta_r} \Theta(p_x) \, .
\end{equation} 
As expected, the total current in the strip is proportional to the difference $\zeta_r - \zeta_l$.  

\begin{figure}[htpb]
\includegraphics[width=0.48\textwidth]{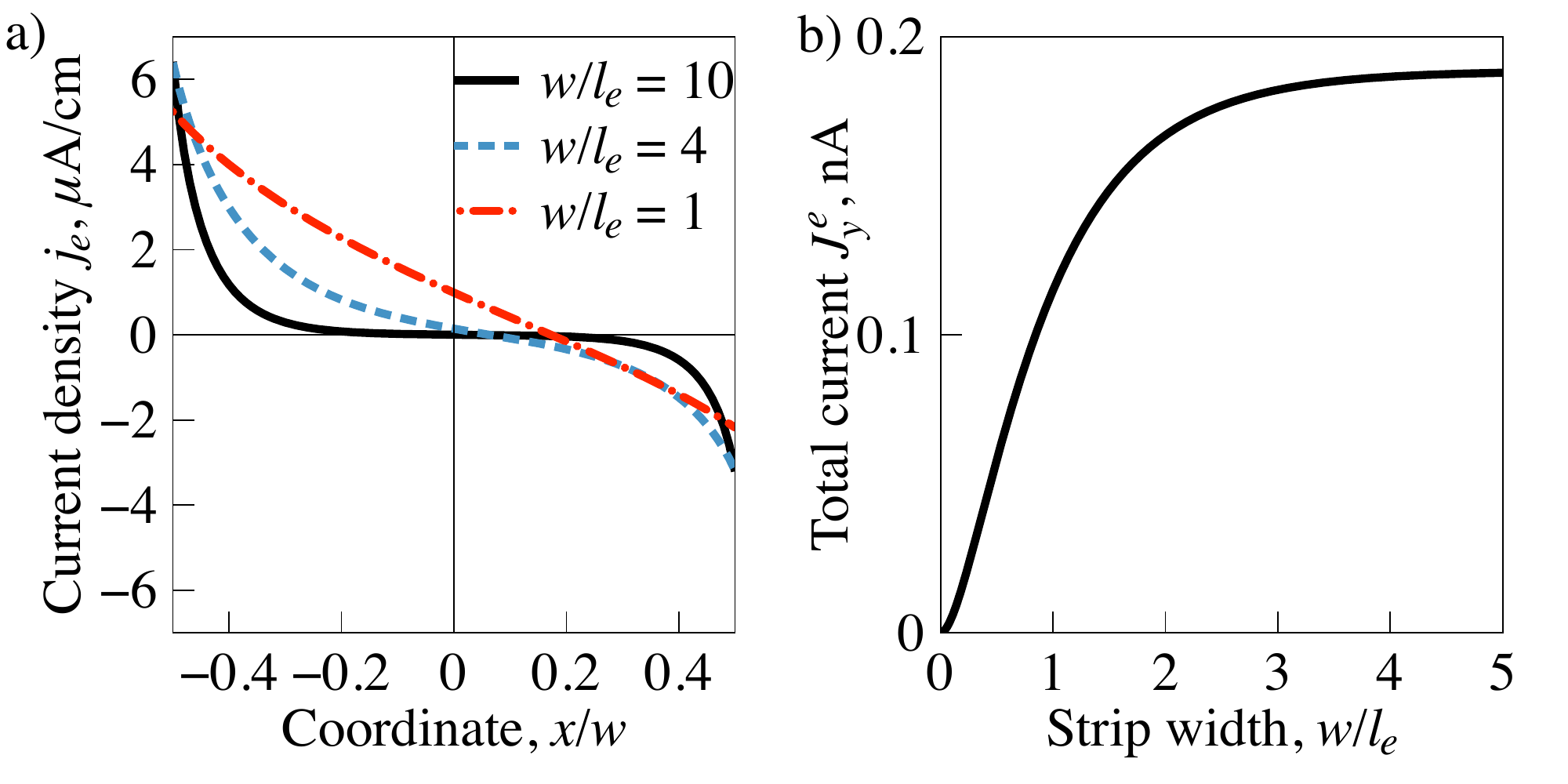}
\caption{\label{fig:fig6} 
(a) Spatial distribution of the electron current density in the strip, where electrons are scattered \textit{specularly} at the left edge ($\zeta_l = 1$) and \textit{diffusely} at the right edge ($\zeta_r = 0$). The distributions are plotted for different ratios between the strip width $w$ and the mean free path $l_e$. (b) Total electron current flowing in the strip as a function of the strip width. The curves are calculated for the same parameters as used in Fig.~\ref{fig:fig3}a.
}
\end{figure} 

Figure~\ref{fig:fig6}a shows the distribution of the electron photocurrent $j_y^e(x)$ in the cross section of the strip. The distributions are
calculated by numeric summation of Eq.~\eqref{j_y_def} with the distribution function given by Eq.~\eqref{f_strip}.  It is assumed that the strip
edges have different scattering properties: Electrons are reflected \textit{specularly} at the left edge ($\zeta_l = 1$) and \textit{diffusely} at the right edge ($\zeta_r = 0$). When the strip width $w$ is much larger than the electron mean free path $l_e$, the photocurrents are concentrated at the edges and vanish in the bulk of the strip. The current densities in that case coincide with those calculated in Sec.~\ref{sec_zero_B} for a semi-infinite structure. The currents at the opposite sides of the strip flow in the opposite directions. The total current at the left edge is twice larger than the current at the right edge, in agreement with Eq.~\eqref{J_electron}. In narrow strips, when the strip width approaches the mean free path, the edge currents merge and the photocurrent is generated in the whole cross section of the strip.

The total photocurrent $J_y^e$ generated in the strip as a function of the strip width is shown in Fig.~\ref{fig:fig6}b. The photocurrent increases with the strip width $w$ in narrow strips and saturates at $w \approx 3l_e$. Further increase of the strip width does not affect the current magnitude.
Projecting the results to $3 \times 3$ mm$^2$ sample with $N = 10^3$ strips we estimate the total current of about $1~\mu$A per W/cm$^2$ in the terahertz spectral range. Such structures based on 2D Dirac materials can be used as fast detectors of terahertz and infrared radiation and its polarization.

\section{Summary} \label{summary}

To summarize, we have studied theoretically the edge photogalvanic effect in 2D materials caused by direct interband optical transitions. The dc electric current emerges due to the optical alignment of electron and hole momenta by linearly polarized electromagnetic wave followed by scattering of  carriers at the edge. The current is formed in a narrow strip near the edge with the width defined by the mean free path at zero magnetic field, and by the diameter of cyclotron orbit at large magnetic fields. At zero magnetic field the edge photocurrent behaves as $\sin 2 \alpha$, where $\alpha$ is the angle between the electric field of the wave and the edge. The current contains counterflowing electron and hole contributions with different magnitudes due to electron-hole asymmetry. 
The excitation spectrum of the edge current calculated for short-range scatterers shows that the current magnitude is larger in gapless materials and reaches 1 nA per W/cm$^2$. Under application of a static magnetic field normal to the sample plane, the edge current is also  excited by the electric field parallel or perpendicular to the edge, and the net current is nonzero even in structures possessing electron-hole symmetry. In a narrow strip made of 2D material, the photocurrent is generated in the whole cross section of the strip and its value integrated over the strip is nonzero in the strip with asymmetric shape or asymmetric static potential. The maximum value of photocurrent is already reached in strips with the width of several mean free paths, which allows one to use a ratchet-like multi-strip structure for the significant increase of the total photocurrent. Our estimations show that the total current can reach 1 $\mu$A per W/cm$^2$ in a 3$\times$3 mm$^2$ sample. Such ratchet-like structures can be used as fast detectors of terahertz and infrared radiation and its polarization.

\acknowledgements

We acknowledge financial support from the Russian Science Foundation (project 17-12-01265). M.V.D. also acknowledges the support from the Basis Foundation for the Advancement of Theoretical Physics and Mathematics and the Russian Federation President Grant No. MK- 2943.2019.2.

\section{Appendix}

Equation for the electric current~\eqref{J_electron} can be also obtained without the explicit calculation of the distribution function~\eqref{f_asym}. To directly calculate $J_y$, we multiply Eq.~\eqref{kinetic_eq} by $e v_y$, sum up the result over $\bm p$ and integrate by $x$, and take into account that
$\sum_{\bm p} v_y g_{\bm p} = 0$ since the optical transitions do not induces a bulk electric current. This yields 
\begin{equation}\label{J_electron2}
J_y^{e} =  - e \nu \sum \limits_{\bm p} \tau_1 v_x v_y [f(\infty,\bm p) - f(0,\bm p)] \,,
\end{equation}
where $\tau_1$ is the relaxation time of the first angular harmonics of the distribution function (also called momentum relaxation time)
defined by $\langle \bm v {\rm St} f \rangle = - \langle \bm v f \rangle /\tau_1$. The distribution function far from the edge can be readily found from the kinetic equation~\eqref{kinetic_eq} with the first term being neglected, which gives $\langle v_x v_y f(\infty , \bm p) \rangle = \tau_2 \langle v_x v_y g_{\bm p} \rangle$, where $\tau_2$ is the relaxation time of the second angular harmonics of the distribution function, 
$\langle v_x v_y {\rm St} f \rangle = - \langle v_x v_y f \rangle /\tau_2$.  For {\it specular} electron reflection from the edge ($\zeta = 1$), the term 
$\sum \limits_{\bm p} \tau_1 v_x v_y f(0,\bm p)$ vanishes,  Eq.~\eqref{J_electron2} coincides with Eq.~\eqref{J_electron} and specifies 
that $\tau^2$ is $\tau_1 \tau_2$.

\end{document}